\documentstyle[twoside,fleqn,espcrc2]{article}
\input prepictex
\input pictex
\input postpictex

\newdimen\xposition
\newdimen\yposition
\newdimen\dyposition
\newdimen\crossbarlength

\def\putcircbar at #1 #2 with fuzz #3 {%
  \xposition=\Xdistance{#1}
  \yposition=\Ydistance{#2}
  \dyposition=\Ydistance{#3}

\setdimensionmode
\put {\Large $\circ$} at {\xposition} {\yposition}

\dimen0 = \yposition
  \advance \dimen0 by -\dyposition
\dimen2 = \yposition
  \advance \dimen2  by \dyposition
\putrule from {\xposition} {\dimen0}
  to {\xposition} {\dimen2}

\dimen4 = \xposition
  \advance \dimen4 by -.5\crossbarlength
\dimen6 = \xposition
  \advance \dimen6 by  .5\crossbarlength
\putrule from {\dimen4} {\dimen0} to {\dimen6} {\dimen0}
\putrule from {\dimen4} {\dimen2} to {\dimen6} {\dimen2}
\setcoordinatemode}

\newdimen\xposition
\newdimen\yposition
\newdimen\dyposition
\newdimen\crossbarlength

\def\putdiamondbar at #1 #2 with fuzz #3 {%
  \xposition=\Xdistance{#1}
  \yposition=\Ydistance{#2}
  \dyposition=\Ydistance{#3}

\setdimensionmode
\put {\Large $\diamond$} at {\xposition} {\yposition}

\dimen0 = \yposition
  \advance \dimen0 by -\dyposition
\dimen2 = \yposition
  \advance \dimen2  by \dyposition
\putrule from {\xposition} {\dimen0}
  to {\xposition} {\dimen2}

\dimen4 = \xposition
  \advance \dimen4 by -.5\crossbarlength
\dimen6 = \xposition
  \advance \dimen6 by  .5\crossbarlength
\putrule from {\dimen4} {\dimen0} to {\dimen6} {\dimen0}
\putrule from {\dimen4} {\dimen2} to {\dimen6} {\dimen2}
\setcoordinatemode}

\newdimen\xposition
\newdimen\yposition
\newdimen\dyposition
\newdimen\crossbarlength

\def\putbigtriangledownbar at #1 #2 with fuzz #3 {%
  \xposition=\Xdistance{#1}
  \yposition=\Ydistance{#2}
  \dyposition=\Ydistance{#3}

\setdimensionmode
\put {$\bigtriangledown$} at {\xposition} {\yposition}

\dimen0 = \yposition
  \advance \dimen0 by -\dyposition
\dimen2 = \yposition
  \advance \dimen2  by \dyposition
\putrule from {\xposition} {\dimen0}
  to {\xposition} {\dimen2}

\dimen4 = \xposition
  \advance \dimen4 by -.5\crossbarlength
\dimen6 = \xposition
  \advance \dimen6 by  .5\crossbarlength
\putrule from {\dimen4} {\dimen0} to {\dimen6} {\dimen0}
\putrule from {\dimen4} {\dimen2} to {\dimen6} {\dimen2}
\setcoordinatemode}
                                           
\newdimen\xposition
\newdimen\yposition
\newdimen\dyposition
\newdimen\crossbarlength

\def\puttrianglebar at #1 #2 with fuzz #3 {%
  \xposition=\Xdistance{#1}
  \yposition=\Ydistance{#2}
  \dyposition=\Ydistance{#3}

\setdimensionmode
\put {$\triangle$} at {\xposition} {\yposition}

\dimen0 = \yposition
  \advance \dimen0 by -\dyposition
\dimen2 = \yposition
  \advance \dimen2  by \dyposition
\putrule from {\xposition} {\dimen0}
  to {\xposition} {\dimen2}

\dimen4 = \xposition
  \advance \dimen4 by -.5\crossbarlength
\dimen6 = \xposition
  \advance \dimen6 by  .5\crossbarlength
\putrule from {\dimen4} {\dimen0} to {\dimen6} {\dimen0}
\putrule from {\dimen4} {\dimen2} to {\dimen6} {\dimen2}
\setcoordinatemode}
                                          
\newdimen\xposition
\newdimen\yposition
\newdimen\dyposition
\newdimen\crossbarlength

\def\puttrianglerightbar at #1 #2 with fuzz #3 {%
  \xposition=\Xdistance{#1}
  \yposition=\Ydistance{#2}
  \dyposition=\Ydistance{#3}

\setdimensionmode
\put {\Large $\triangleright$} at {\xposition} {\yposition}

\dimen0 = \yposition
  \advance \dimen0 by -\dyposition
\dimen2 = \yposition
  \advance \dimen2  by \dyposition
\putrule from {\xposition} {\dimen0}
  to {\xposition} {\dimen2}

\dimen4 = \xposition
  \advance \dimen4 by -.5\crossbarlength
\dimen6 = \xposition
  \advance \dimen6 by  .5\crossbarlength
\putrule from {\dimen4} {\dimen0} to {\dimen6} {\dimen0}
\putrule from {\dimen4} {\dimen2} to {\dimen6} {\dimen2}
\setcoordinatemode}
                        
\newdimen\xposition
\newdimen\yposition
\newdimen\dyposition

\def\puttriangleleftbar at #1 #2 with fuzz #3 {%
  \xposition=\Xdistance{#1}
  \yposition=\Ydistance{#2}
  \dyposition=\Ydistance{#3}

\setdimensionmode
\put {\Large $\triangleleft$} at {\xposition} {\yposition}

\dimen0 = \yposition
  \advance \dimen0 by -\dyposition
\dimen2 = \yposition
  \advance \dimen2  by \dyposition
\putrule from {\xposition} {\dimen0}
  to {\xposition} {\dimen2}

\dimen4 = \xposition
  \advance \dimen4 by -.5\crossbarlength
\dimen6 = \xposition
  \advance \dimen6 by  .5\crossbarlength
\putrule from {\dimen4} {\dimen0} to {\dimen6} {\dimen0}
\putrule from {\dimen4} {\dimen2} to {\dimen6} {\dimen2}
\setcoordinatemode}

\newcommand{\AmS}{{\protect\the\textfont2
  A\kern-.1667em\lower.5ex\hbox{M}\kern-.125emS}}

\title{Better Actions\thanks{Research supported by the
U.~S. Department of Energy under
contract DE-FG03-92ER40732/B004}}

\author{K.~Cahill\thanks{E-mail: kevin @ cahill.phys.unm.edu}
and G.~Herling\thanks{Member of the Center for Advanced Studies;
e-mail: herling@unm.edu}\\
{ \hskip 1cm } \\
New Mexico Center for Particle Physics, University of New Mexico,
Albuquerque, NM 87131-1156, USA\\
Division de Physique Th\'eorique,\thanks
{Unit\'e de Recherche des Universit\'es Paris XI 
et Paris VI associ\'ee au CNRS.} 
Institut de Physique Nucl\'eaire,
91406 Orsay Cedex, France}

\begin{document}

\begin{abstract}
We explain why compact $ U(1) $ confines and how to fix it.
We show that plaquettes of negative trace 
carry most of the confinement signal in compact $ SU(2) $\null.
We show how to perform noncompact gauge-invariant
simulations without auxiliary fields.
We suggest a way to simulate fermions 
without doublers.
\end{abstract}

\maketitle

\section{INTRODUCTION}

We reort here the results
of four different studies
all of which are attempts
to find better lattice actions.
The first study explains why
compact $U(1)$ gauge theory displays
confinement at strong coupling and
shows how to remove this artifact.
The second study shows that most
of the confinement signal in compact $SU(2)$ gauge theory  
is due to plaquettes of negative trace.
These results obtain for
both the Wilson action and the Manton action.
The third study shows how to 
perform noncompact lattice simulations
that are exactly gauge invariant and that
do not involve auxiliary fields.  
The fourth study presents a lattice action
for fermions that avoids doublers.

\section{WHY COMPACT $ U(1) $ CONFINES}

Compact lattice simulations of $ U(1) $ gauge theory
display confinement at strong coupling,
as shown by Figure 1 which plots
Creutz ratios\cite{Creu80} obtained with the Wilson action.
This lattice artifact is obvious
in the figure at $ \beta = 0.25 $
and at $ \beta = 0.5 $,
and incipient at $ \beta = 1 $.
It arises because $ U(1) $ is a circle;
if one cuts the circle, then there is no confinement,
as shown in the figure by the $\chi(i,j)$'s
labeled ``cut {\footnotesize $\bigcirc$}''
which follow the curves
of the exact Creutz ratios down
to $ \beta = 0.25 $. 

\begin{figure}
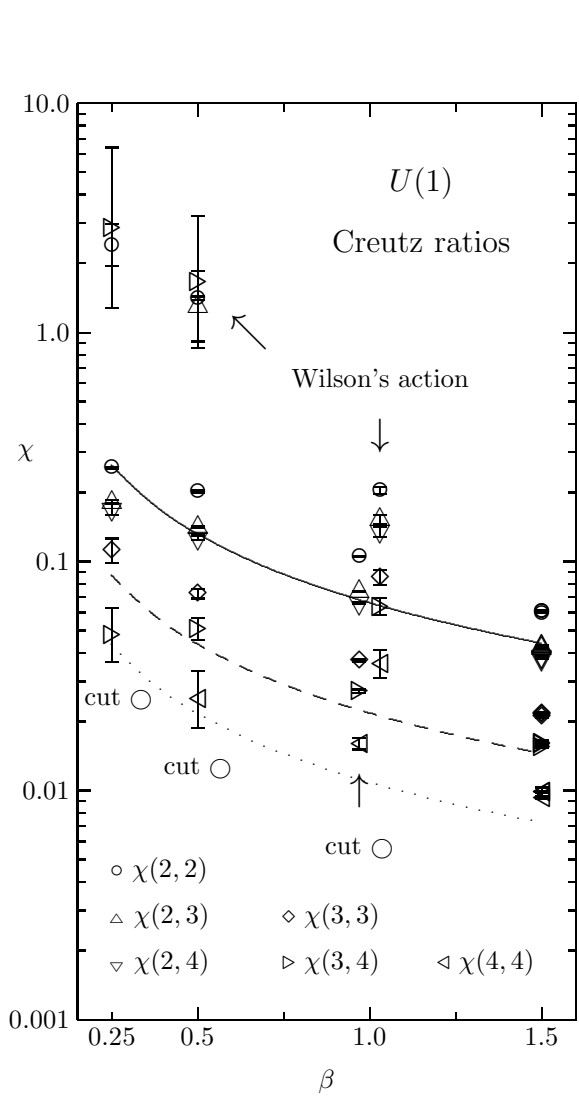
   
\beginpicture
\crossbarlength=5pt

\ticksin
\inboundscheckon
\setcoordinatesystem units <1.80in,1.20in>
\setplotarea x from 0.15 to 1.6, y from -3.0 to 1.0
\axis bottom label {$\beta$} ticks
  numbered at 0.5 1.0 1.5 /
  numbered short at 0.25 /
  unlabeled short from 0.75 to 0.75 by 0.25
  from 1.25 to 1.25 by 0.25 /
\axis left  
  ticks logged
  numbered at 0.001 0.01 0.1 1.0 10.0 /
  unlabeled short at 1.0 /
  withvalues {$\chi$ \quad} /
  at 0.3 /
  unlabeled short from 2.0 to 9.0 by 1.0
  from 0.2 to 0.9 by 0.1
  from 0.02 to 0.09 by 0.01
  from 0.002 to 0.009 by 0.001 /
\axis top label {}
  ticks
  unlabeled at 0.5 1.0 1.5 /
  unlabeled short from 0.25 to 1.25 by 0.25 /
\axis right ticks logged
  unlabeled at 0.001 0.01 0.1 1.0 10.0 /
  unlabeled short from 2.0 to 9.0 by 1.0
  from 0.2 to 0.9 by 0.1
  from 0.02 to 0.09 by 0.01
  from 0.002 to 0.009 by 0.001 /

\put {\large $U(1)$} at 1.15 0.65
\put {\large Creutz ratios} at 1.15 .40
\put {Wilson's action} at 1.03 -0.20
\put {$\circ$  $\chi(2,2)$ } at 0.4 -2.35
\put {{\tiny $\triangle$}  $\chi(2,3)$ } at 0.4 -2.55
\put {{\tiny $\bigtriangledown$}  $\chi(2,4)$ } at 0.4 -2.75
\put {$\diamond$  $\chi(3,3)$ } at 0.9 -2.55
\put {$\triangleright$  $\chi(3,4)$ } at 0.9 -2.75
\put {$\triangleleft$  $\chi(4,4)$ } at 1.35 -2.75
\put {cut \footnotesize $\bigcirc$} at 0.27 -1.6
\put {cut \footnotesize $\bigcirc$} at 0.50 -1.9
\put {cut \footnotesize $\bigcirc$} at 0.97 -2.25
\put {\Large $\uparrow$} at 0.97 -2.00
\put {\Large $\nwarrow$} at 0.65 0.00
\put {\Large $\downarrow$} at 1.03 -0.45
                                                                 
\setquadratic
\plot 0.25 -0.57787 0.375 -0.75396 0.5 -0.87890 0.625 -0.97581
0.75 -1.05499 0.875 -1.12194 1.0 -1.1799 1.125 -1.23108
1.25 -1.27684 1.375 -1.31823 1.5 -1.3561 /  
\setdashes
\plot 0.25 -1.05909 0.375 -1.23518 0.5 -1.36012 0.625 -1.45703
0.75 -1.53621 0.875 -1.60316 1.0 -1.6611 1.125 -1.71230
1.25 -1.75806 1.375 -1.79945 1.5 -1.8371 /
\setdots
\setquadratic
\plot 0.25 -1.36123 0.375 -1.53732 0.5 -1.66226 0.625 -1.75917
0.75 -1.83835 0.875 -1.90530 1.0 -1.9634 1.125 -2.01444
1.25 -2.06020 1.375 -2.10159 1.5 -2.1397 /  
\setsolid

\inboundscheckon

\putcircbar at     0.250  0.380801 with fuzz  0.091262
\puttrianglerightbar at     0.250  0.457295 with fuzz  0.349989

\putcircbar at     0.500  0.149231 with fuzz  0.008010
\puttrianglebar at     0.500  0.113528 with fuzz  0.152936
\puttrianglerightbar at     0.500  0.218946 with fuzz  0.287882

\putcircbar at     1.030 -0.690748 with fuzz  0.016234
\puttrianglebar at     1.030 -0.815917 with fuzz  0.020200
\putdiamondbar at     1.030 -1.068452 with fuzz  0.032026
\putbigtriangledownbar at     1.030 -0.869624 with fuzz  0.022118
\puttrianglerightbar at     1.030 -1.196638 with fuzz  0.037508
\puttriangleleftbar at     1.030 -1.447267 with fuzz  0.060566

\putcircbar at     1.500 -1.218008 with fuzz  0.000368
\puttrianglebar at     1.500 -1.370503 with fuzz  0.000586
\putdiamondbar at     1.500 -1.666935 with fuzz  0.005745
\putbigtriangledownbar at     1.500 -1.415428 with fuzz  0.002895
\puttrianglerightbar at     1.500 -1.798606 with fuzz  0.009116
\puttriangleleftbar at     1.500 -2.005925 with fuzz  0.002696


\putcircbar at     0.250 -0.589819 with fuzz  0.003888
\puttrianglebar at     0.250 -0.739796 with fuzz  0.010155
\putdiamondbar at     0.250 -0.953098 with fuzz  0.053898
\putbigtriangledownbar at     0.250 -0.772305 with fuzz  0.026694
\puttrianglerightbar at     0.250 -1.322011 with fuzz  0.118639

\putcircbar at     0.500 -0.694922 with fuzz  0.004533
\puttrianglebar at     0.500 -0.849077 with fuzz  0.004173
\putdiamondbar at     0.500 -1.140439 with fuzz  0.019775
\putbigtriangledownbar at     0.500 -0.896303 with fuzz  0.010157
\puttrianglerightbar at     0.500 -1.293009 with fuzz  0.047972
\puttriangleleftbar at     0.500 -1.602520 with fuzz  0.123682

\putcircbar at     0.97 -0.976836 with fuzz  0.000825
\puttrianglebar at     0.97 -1.131287 with fuzz  0.001624
\putdiamondbar at     0.97 -1.431269 with fuzz  0.004387
\putbigtriangledownbar at     0.97 -1.179616 with fuzz  0.002910
\puttrianglerightbar at     0.97 -1.566081 with fuzz  0.007517
\puttriangleleftbar at     0.97 -1.795094 with fuzz  0.025128

\putcircbar at     1.500 -1.216392 with fuzz  0.002038
\puttrianglebar at     1.500 -1.370034 with fuzz  0.003157
\putdiamondbar at     1.500 -1.663625 with fuzz  0.006889
\putbigtriangledownbar at     1.500 -1.417037 with fuzz  0.004180
\puttrianglerightbar at     1.500 -1.790585 with fuzz  0.011757
\puttriangleleftbar at     1.500 -2.005431 with fuzz  0.019894

\putcircbar at     1.500 -1.224934 with fuzz  0.000699
\puttrianglebar at     1.500 -1.378484 with fuzz  0.001239
\putdiamondbar at     1.500 -1.678268 with fuzz  0.004380
\putbigtriangledownbar at     1.500 -1.425784 with fuzz  0.000191
\puttrianglerightbar at     1.500 -1.810215 with fuzz  0.003361
\puttriangleleftbar at     1.500 -2.031186 with fuzz  0.004726
                                                                        
\endpicture
\caption{The $U(1)\/$ Creutz ratios $\chi(i,j)$
as given by Wilson's action 
and by Wilson's action on a cut circle.
The curves represent the exact Creutz ratios,
$\chi(2,2)$ (solid), $\chi(3,3)$ (dashes), and $\chi(4,4)$ (dots).
At $\beta = 1$, the symbols for the Wilson action
are plotted to the right of those for the cut circle;
at $\beta = 1.5$ they overlap.
Wilson's action on the full circle
confines for $ \beta < 1 $.}
\end{figure}

\par
We ran on a $ 12^4 $ lattice
and began all runs from a cold start
in which all links were unity.
We used a Metropolis algorithm 
and rejected any plaquette whose 
phase $ \theta $ was either 
greater than $ \pi - \epsilon $
or less than $ - \pi + \epsilon $.
We saw no confinement signal as long as the step size
was smaller than the 
thickness $ 2 \epsilon $ 
of the cut in the phase $ \theta $.
Simulations with Manton's action
show similar results.
\par
One may interpret these results in terms
of monopoles.  Since the phase $\theta$ of each
plaquette is required to lie between $ \pi - \epsilon $
and $ - \pi + \epsilon $,
it follows that for $ \epsilon > 0 $,
no string can ever penetrate any plaquette.
For the Wilson action we took $ .02 < \epsilon < 0.1 $
and noticed no sensitivity to $ \epsilon $
within that range.  For the Manton action,
we took $ \epsilon = 0.026 $ in all runs with cut circles.

\section{COMPACT $ SU(2) $}

\par
In view of these results for $U(1)$,
one might wonder whether similar lattice
artifacts exist in the case of the group $SU(2)$.
Inasmuch as $U(1)$ and $SU(2)$ have different
first homotopy groups ($ \pi_1( U(1) ) = Z $ but
$ \pi_1( SU(2) ) = 0 $),
one might assume that excising a small cap 
around the antipode ($ g = -1 $) on the $ SU(2) $ group manifold 
(the three sphere $ S_3 $ in four dimensions)
would have little effect on Creutz ratios.
\par
To check this assumption,
we ran from cold starts
on an $ 8^4 $ lattice
and used a Metropolis algorithm
in which we rejected plaquettes
that lay within a small cap around the antipode.
The step size was small compared to 
the size of the excluded cap.
As shown in Figure 2,
the Creutz ratios $ \chi(i,j) $
do not depend upon whether the small cap
was excluded.
\par
But what about the excision of a large cap?
To study this question,
we again began with cold starts
on an $ 8^4 $ lattice
and employed a Metropolis algorithm
with a small step size.
We rejected all plaquettes that had a negative trace, 
thus excluding half of the $ SU(2) $ sphere.

\begin{figure} [t]
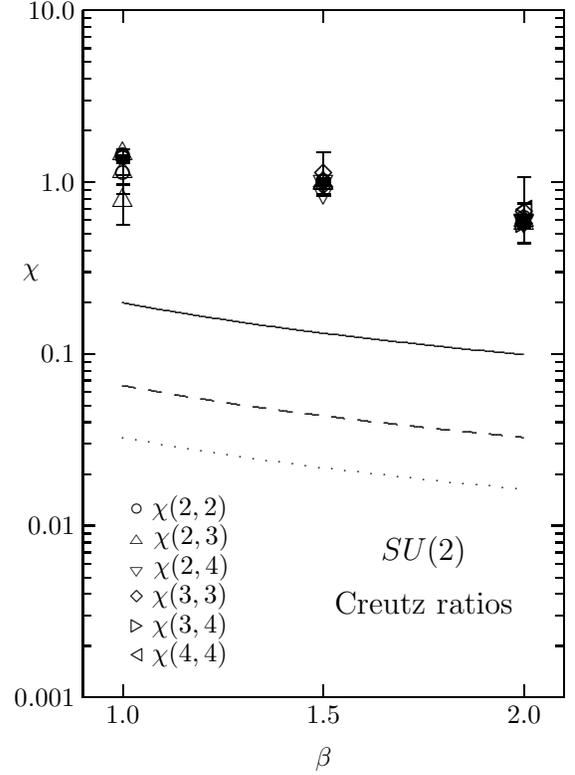
  
\beginpicture
\crossbarlength=5pt

\ticksin
\inboundscheckon
\setcoordinatesystem units <2.10in,0.90in>   
\setplotarea x from 0.90 to 2.1, y from -3.0 to 1.0
\axis bottom label {$\beta$} ticks
  numbered at 1.0 1.5 2.0 / /
\axis left 
  ticks logged 
  numbered at 0.001 0.01 0.1 1.0 10.0 /
  unlabeled short at 1.0 /
  withvalues {$\chi$ \quad} /
  at 0.3 /
  unlabeled short from 2.0 to 9.0 by 1.0
  from 0.2 to 0.9 by 0.1
  from 0.02 to 0.09 by 0.01 
  from 0.002 to 0.009 by 0.001 /
\axis top label {}
  ticks
  unlabeled at 1.0 1.5 2.0 / /
\axis right ticks logged
  unlabeled at 0.001 0.01 0.1 1.0 10.0 /
  unlabeled short from 2.0 to 9.0 by 1.0
  from 0.2 to 0.9 by 0.1
  from 0.02 to 0.09 by 0.01 
  from 0.002 to 0.009 by 0.001 /

\put {\large $SU(2)$} at 1.75 -2.15
\put {\large Creutz ratios} at 1.75 -2.45
\put {$\circ$  $\chi(2,2)$ } at 1.15 -1.90
\put {{\tiny $\triangle$}  $\chi(2,3)$ } at 1.15 -2.07
\put {{\tiny $\bigtriangledown$}  $\chi(2,4)$ } at 1.15 -2.24
\put {$\diamond$  $\chi(3,3)$ } at 1.15 -2.41
\put {$\triangleright$  $\chi(3,4)$ } at 1.15 -2.58
\put {$\triangleleft$  $\chi(4,4)$ } at 1.15 -2.75

\setquadratic
\plot 1.0 -0.70281011 1.25 -0.79972012
1.5 -0.87890137 1.75 -0.94584816 2.0 -1.00384010 /
\setdashes
\plot 1.0 -1.18402699 1.25 -1.28093700
1.5 -1.36011825 1.75 -1.42706504 2.0 -1.48505698 /
\setdots
\setquadratic
\plot 1.0 -1.48616962 1.25 -1.58307964
1.5 -1.66226088 1.75 -1.72920767 2.0 -1.78719962 /
\setsolid

\inboundscheckon


\putcircbar at  1.000  0.153134 with fuzz  0.002580
\puttrianglebar at  1.000  0.166729 with fuzz  0.023579

\putcircbar at  1.500  0.001729 with fuzz  0.000622
\puttrianglebar at  1.500 -0.009273 with fuzz  0.002756
\putdiamondbar at  1.500  0.053458 with fuzz  0.120702
\putbigtriangledownbar at  1.500 -0.059036 with fuzz  0.019513

\putcircbar at  2.000 -0.210732 with fuzz  0.000272
\puttrianglebar at  2.000 -0.223445 with fuzz  0.000510
\putdiamondbar at  2.000 -0.243467 with fuzz  0.005434
\putbigtriangledownbar at  2.000 -0.226618 with fuzz  0.002153
\puttrianglerightbar at  2.000 -0.245016 with fuzz  0.021871
\puttriangleleftbar at  2.000 -0.162612 with fuzz  0.192509


\putcircbar at  1.000  0.145263 with fuzz  0.013535
\puttrianglebar at  1.000 -0.104510 with fuzz  0.144819

\putcircbar at  1.000  0.139237 with fuzz  0.017328
\puttrianglebar at  1.000  0.061568 with fuzz  0.130566

\putcircbar at  1.000  0.048537 with fuzz  0.062059

\putcircbar at  1.500 -0.002333 with fuzz  0.000411
\puttrianglebar at  1.500 -0.002328 with fuzz  0.002040
\putdiamondbar at  1.500 -0.027499 with fuzz  0.052805
\putbigtriangledownbar at  1.500  0.003521 with fuzz  0.016341

\putcircbar at  2.000 -0.227384 with fuzz  0.002856
\puttrianglebar at  2.000 -0.240667 with fuzz  0.009427
\putdiamondbar at  2.000 -0.177931 with fuzz  0.053960
\putbigtriangledownbar at  2.000 -0.221281 with fuzz  0.027810

\endpicture
\caption{The $SU(2)\/$ Creutz ratios $\chi(i,j)$
as given by Wilson's action
with and without the exclusion
of a small part of the $ SU(2) $ sphere.
The curves represent the perturbative Creutz ratios,
$\chi(2,2)$ (solid), $\chi(3,3)$ (dashes), and $\chi(4,4)$ (dots).
The $\chi(i,j)$'s substantially overlap.}
\end{figure}

\par
In Figure 3 we plot the Creutz ratios
$ \chi(i,j) $ both for the usual Wilson
action and for the positive-plaquette Wilson action. 
As shown in the figure,
the $ \chi(i,j) $'s of the positive-plaquette
simulations exhibit behavior that is 
substantially perturbative. 
The confinement signal has disappeared.
Apparantly the plaquettes of negative trace carry 
most of the confinement signal.

\begin{figure}
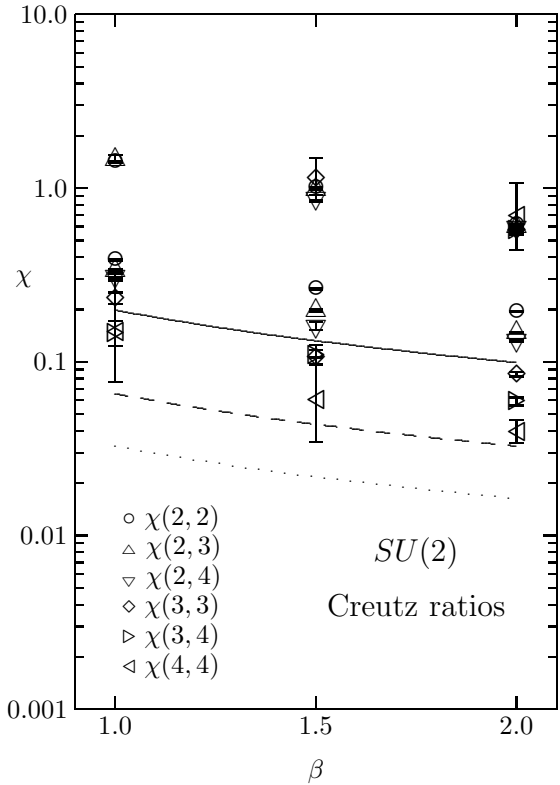
 
\beginpicture
\crossbarlength=5pt

\ticksin
\inboundscheckon
\setcoordinatesystem units <2.10in,0.91in>   
\setplotarea x from 0.90 to 2.1, y from -3.0 to 1.0
\axis bottom label {$\beta$} ticks
  numbered at 1.0 1.5 2.0 / /
\axis left 
  ticks logged 
  numbered at 0.001 0.01 0.1 1.0 10.0 /
  unlabeled short at 1.0 /
  withvalues {$\chi$ \quad} /
  at 0.3 /
  unlabeled short from 2.0 to 9.0 by 1.0
  from 0.2 to 0.9 by 0.1
  from 0.02 to 0.09 by 0.01 
  from 0.002 to 0.009 by 0.001 /
\axis top label {}
  ticks
  unlabeled at 1.0 1.5 2.0 / /
\axis right ticks logged
  unlabeled at 0.001 0.01 0.1 1.0 10.0 /
  unlabeled short from 2.0 to 9.0 by 1.0
  from 0.2 to 0.9 by 0.1
  from 0.02 to 0.09 by 0.01 
  from 0.002 to 0.009 by 0.001 /

\put {\large $SU(2)$} at 1.75 -2.1
\put {\large Creutz ratios} at 1.75 -2.4
\put {$\circ$  $\chi(2,2)$ } at 1.15 -1.90
\put {{\tiny $\triangle$}  $\chi(2,3)$ } at 1.15 -2.07
\put {{\tiny $\bigtriangledown$}  $\chi(2,4)$ } at 1.15 -2.24
\put {$\diamond$  $\chi(3,3)$ } at 1.15 -2.41
\put {$\triangleright$  $\chi(3,4)$ } at 1.15 -2.58
\put {$\triangleleft$  $\chi(4,4)$ } at 1.15 -2.75

\setquadratic
\plot 1.0 -0.70281011 1.25 -0.79972012
1.5 -0.87890137 1.75 -0.94584816 2.0 -1.00384010 /
\setdashes
\plot 1.0 -1.18402699 1.25 -1.28093700
1.5 -1.36011825 1.75 -1.42706504 2.0 -1.48505698 /
\setdots
\setquadratic
\plot 1.0 -1.48616962 1.25 -1.58307964
1.5 -1.66226088 1.75 -1.72920767 2.0 -1.78719962 /
\setsolid

\inboundscheckon


\putcircbar at  1.000  0.153134 with fuzz  0.002580
\puttrianglebar at  1.000  0.166729 with fuzz  0.023579

\putcircbar at  1.500  0.001729 with fuzz  0.000622
\puttrianglebar at  1.500 -0.009273 with fuzz  0.002756
\putdiamondbar at  1.500  0.053458 with fuzz  0.120702
\putbigtriangledownbar at  1.500 -0.059036 with fuzz  0.019513

\putcircbar at  2.000 -0.210732 with fuzz  0.000272
\puttrianglebar at  2.000 -0.223445 with fuzz  0.000510
\putdiamondbar at  2.000 -0.243467 with fuzz  0.005434
\putbigtriangledownbar at  2.000 -0.226618 with fuzz  0.002153
\puttrianglerightbar at  2.000 -0.245016 with fuzz  0.021871
\puttriangleleftbar at  2.000 -0.162612 with fuzz  0.192509


\putcircbar at  1.000 -0.412618 with fuzz  0.004624
\puttrianglebar at  1.000 -0.473873 with fuzz  0.007737
\putdiamondbar at  1.000 -0.634087 with fuzz  0.032932
\putbigtriangledownbar at  1.000 -0.508353 with fuzz  0.012939
\puttrianglerightbar at  1.000 -0.837058 with fuzz  0.071378
\puttriangleleftbar at  1.000 -0.826819 with fuzz  0.292081

\putcircbar at  1.500 -0.579097 with fuzz  0.004957
\puttrianglebar at  1.500 -0.702691 with fuzz  0.007676
\putdiamondbar at  1.500 -0.973854 with fuzz  0.041195
\putbigtriangledownbar at  1.500 -0.794371 with fuzz  0.021743
\puttrianglerightbar at  1.500 -0.958810 with fuzz  0.054138
\puttriangleleftbar at  1.500 -1.219334 with fuzz  0.243334

\putcircbar at  2.000 -0.711866 with fuzz  0.002184
\puttrianglebar at  2.000 -0.831365 with fuzz  0.003742
\putdiamondbar at  2.000 -1.072320 with fuzz  0.012578
\putbigtriangledownbar at  2.000 -0.879215 with fuzz  0.004578
\puttrianglerightbar at  2.000 -1.228597 with fuzz  0.023080
\puttriangleleftbar at  2.000 -1.402292 with fuzz  0.065563

\endpicture
\caption{The $SU(2)\/$ Creutz ratios $\chi(i,j)$
as given by Wilson's action
with and without the exclusion
of half the $ SU(2) $ sphere.
The curves represent the perturbative Creutz ratios,
$\chi(2,2)$ (solid), $\chi(3,3)$ (dashes), and $\chi(4,4)$ (dots).
In the two cases,
the $\chi(i,j)$'s substantially differ.}
\end{figure}

\section{NONCOMPACT SIMULATIONS}

Suppose we write the fermion field
\begin{equation}
\psi = \pmatrix{\psi_1\cr \psi_2\cr \vdots\cr
\psi_n\cr}
\label{psi=}
\end{equation}
in terms of some orthonormal basis vectors $e_a(x)$
that may vary
with position and time
\begin{equation}
\psi(x) = \psi_a(x) e_a(x).
\label{psi(x)=}
\end{equation}
Then the derivative $\partial_\mu \psi(x)$ has two terms:
\begin{equation}
\partial_\mu \psi(x) =
e_a(x) \partial_\mu \psi_a(x) + \psi_a(x) \partial_\mu e_a(x).
\end{equation}
And so if we let the gauge field $A^{ab}_\mu(x)$ be
\begin{equation}
A^{ab}_\mu(x) = (-i/g) e^\dagger_a(x) \cdot \partial_\mu e_{b}(x),
\label{A=}
\end{equation}
then the free action density 
$ i \bar \psi \gamma^\mu \partial_\mu \psi $ 
becomes
\begin{equation}
{\cal L_D} = i \bar \psi_a \gamma^\mu
\left( \delta_{ab} \partial_\mu
+ i A^{ab}_\mu \right) \psi_b
= i \bar \psi_a \gamma^\mu D^{ab}_\mu \psi_b
\end{equation}
as in a gauge theory.
\par
Under a gauge transformation
\begin{equation}
\psi'_a( x ) = g_{ab}( x ) \psi_b( x ),
\end{equation}
the field $ \psi( x ) $ and therefore
the action is invariant if the vectors
$ e_a( x ) $ transform as
\begin{equation}
e'_a( x ) = g^{-1}_{ca}( x ) e_c( x )
\end{equation}
which for unitary groups is
$ e'_a( x ) = g^*_{ac}( x ) e_c( x ) $.
\par
To generate $ U(1) $ gauge fields,
one may use a single normalized
complex three-vector,
\begin{equation}
e( x ) = e^{i\alpha( x )} \pmatrix{ x_1 ( x ) \cr
x_2 ( x ) + i y_2 ( x ) \cr
x_3 ( x ) + i y_3 ( x ) \cr } ,
\label{e=}
\end{equation}
where $ x_1 ( x ) \ge 0 $\null.
For $ U(2), $ one may use two
orthonormal complex five-vectors\cite{vectors}.
\par
We have successfully used this method
to simulate $ U(1) $ and are now applying it to $ SU(2) $.

\section{FERMIONS WITHOUT DOUBLERS}
On the lattice each species
of fermion appears as 16
different fermions.
The root of this problem
is that the natural discretization
of the Fermi action
approximates the derivative
by means of a gap of two lattice spacings.
\par
At the price of some nonlocality,
we may leave out the unwanted states
from the start\cite{doub}.
Thus on a lattice of even size $ N = 2 F $,
we may place independent fermionic variables
$ \psi ( 2n ) $ and $ \overline \psi( 2n ) $
only on the $ F^4 $ even sites $ 2 n $
where $ n $ is a four-vector of integers,
\begin{equation}
2 n = ( 2n_1, 2n_2, 2n_3, 2n_4 )
\label{even}
\end{equation}
and $ 1 \le n_i \le F $ for $ i = 1, .., 4 $\null.
To extend the variables $ \psi ( 2n ) $
and $ \overline \psi( 2n ) $ to the nearest-neighbor
sites $ 2n \pm \hat \mu $, we first define the Fourier variables
$\widetilde{\psi} ( k ) $ (and $ \overline{ \widetilde{\psi} } ( k ) $),
\begin{equation}
\widetilde{\psi} ( k ) =
{1 \over F^2 }
\sum_{n}
\exp\left[ - i { 2 \pi n \cdot k \over F } \right]
\psi( 2n )
\label{tilde psi}
\end{equation}
in which the sum extends over each $ n_i, i=1,\dots,4 $
from 1 to $ F $ and $ k $ is a four-vector
of integers $ k = ( k_1, k_2, k_3, k_4 ) $
with $ 1 \le k_i \le F $\null.
In terms of these Fourier variables,
the dependent, nearest-neighbor variable $ \psi ( 2n + \hat \mu ) $ is
\begin{equation}
\psi ( 2n + \hat \mu ) = { 1 \over F^2 }
\sum_{k}
e^{ i\pi \left( 2n + \hat \mu \right) \cdot k / F }
\widetilde{\psi} ( k ).
\label{neighbor f}
\end{equation}
in which the sum extends over each $ k_i, i = 1,..,4 $
from 1 to $ F $.
In terms of the lattice delta functions
\begin{equation}
\delta^\pm( 2 j ) = { 1 \over F }
\sum_{ l = 1 }^F
\exp
\left[ i { 2 \pi \left( \pm {1 \over 2 } - j \right)
l \over F } \right],
\label{delta}
\end{equation}
we may write 
the nearest-neighbor variables as the sums
\begin{equation}
\psi ( 2n \pm \hat \mu ) =
\sum_{j = 1 - n_\mu }^{ F - n_\mu  }
\delta^\pm( 2 j ) \psi ( 2n + 2j \hat \mu ).
\label{nn}
\end{equation}
\par
The action with independent fields at only $ F^4 $ sites is
\begin{eqnarray}
S & = & (2a)^4
\sum_{n}
\overline{ \psi } ( 2n )
\big\{
- m \, \psi( 2n )
\nonumber \\
& & \mbox{}
- \sum_{\mu = 1}^4
{ \gamma_\mu \over 2ai }
\left[
\psi( 2n + \hat \mu ) - \psi( 2n - \hat \mu )
\right]
\big\},
\label{action}
\end{eqnarray}
in which the nearest-neighbor variables
$ \psi ( 2n \pm \hat \mu ) $
are given by eq.(\ref{nn})
and the sum is over $ 1 \le n_i \le F $
as in eq.(\ref{tilde psi}).
\par
We may now verify that there are no doublers.
The Fourier series (\ref{tilde psi}) and (\ref{neighbor f})
diagonalize the action (\ref{action}) and 
the lattice propagator is
\begin{equation}
{ - m a + \sum_\mu \gamma_\mu \sin \left( { \pi k_\mu \over F } \right)
\over
 m^2 a^2 + \sum_\mu \sin^2 \left( { \pi k_\mu \over F } \right) }.
\label{prop}
\end{equation}
For $ m = 0 $ this propagator
has poles only at $ k_\mu = F $,
which is the same point as $ k_\mu = 0 $\null.
There are no doublers.
Gauge fields may be
added in the usual way
or in the vectorial manner of section 4.
In a gauge theory with a gauge field $ U_\mu( n ) $
defined on the link $ ( n, n + \hat \mu ) $,
one may construct
the ordered product
$ U( 2n, 2n + 2j \hat \mu ) $
of Wilson links $ U_\mu( n ) $
from site $ 2n + 2j \hat \mu $ to site $ 2n $
for $ j > 0 $.
Thus by inserting the line $ U( 2n, 2n + 2j \hat \mu ) $
into the action (\ref{action}),
one may render it covariant.
This procedure should also work for
chiral gauge theories.
\par
Because of the lack of locality,
the fermion propagator is not as sparse
as the usual propagator.
On the other hand,
there are only one-sixteenth as many
fermionic variables,
and so the fermion propagator
is smaller by a factor of 256.
The present formalism
of thinned fermions
therefore may be useful
in practice as well as in principle.
\par
We intend to test this idea by
simulating QED in two dimensions.

\section*{ACKNOWLEDGMENTS}
\par
We are indebted to
M.~Creutz, G.~Marsaglia, W.~Press, and J.~Smit
for useful conversations,
to the Department of Energy for support
under grant DE-FG03-92ER40732/B004, and to
B.~Dieterle and the Maui Center for High-Performance
Computing\footnote[1]
{Research sponsored in part by the Phillips Laboratory, Air
     Force Materiel Command, USAF, under cooperative agreement
     F29601-93-2-0001.  The U.S. Government retains a
     nonexclusive copyright to this work.  The views and conclusions
     of this work are those of the authors and
     should not be interpreted as necessarily representing the
     official policies or endorsements, either expressed or
     implied, of Phillips Laboratory or the U.S. Government.}
for computer time.

\end{document}